\documentclass{article}
\usepackage{amsmath,amssymb}

\newtheorem{theorem}{Theorem}
\newtheorem{lemma}{Lemma}

\begin{document}

\title{Triple-Error-Correcting BCH-Like Codes}

\author{Carl Bracken$^1$,  Tor Helleseth$^2$\\
$^1$Department of Mathematics, National University of Ireland\\
Maynooth, Co. Kildare, Ireland\\
$^2$Department of Informatics, University of Bergen\\
PB 7803, N-5020 Bergen, Norway}

\maketitle

\begin{abstract}
The binary primitive triple-error-correcting BCH code is a cyclic code of minimum distance $d=7$ with generator polynomial having zeros $\alpha$, $\alpha^3$ and $\alpha^5$ where $\alpha$ is a primitive $(2^n-1)$-root of unity. The zero set of the code is said to be $\{1,3,5\}$. In the 1970's Kasami showed that one can construct similar triple-error-correcting codes using zero sets consisting of different triples than the BCH codes. Furthermore, in 2000 Chang et. al. found new triples leading to triple-error-correcting codes. In this paper a new such triple is presented. In addition a new method is presented that may be of interest in finding further such triples. The method is illustrated by giving a new and simpler proof of one of the known Kasami triples $\{1,2^k+1,2^{3k}+1\}$ where $n$ is odd and $\gcd(k,n)=1$ as well as to find the new triple given by $\{1,2^k+1,2^{2k}+1\}$ for any $n$ where $\gcd(k,n)=1$.
\end{abstract}

\section{Introduction}
 The well known $t$-error-correcting BCH codes were found by Bose and Chaudhuri~\cite{BC} and Hocquenghem~\cite{H} and they have been a topic for thorough investigations. The binary primitive triple-error-correcting BCH code is a cyclic code of minimum distance $d=7$ with generator polynomial $g(x)$ having zeros $\alpha$, $\alpha^3$ and $\alpha^5$ where $\alpha$ is a primitive $(2^n-1)$-root of unit in $GF(2^n)$, the finite field with $2^n$ elements. The zero set of the code is said to be the triple $\{1,3,5\}$. Kasami~\cite{Kas1} and Chang et. al.~\cite{Cha} showed that one can construct similar triple-error-correcting codes using zero sets consisting of different triples. One of the Kasami triples are $\{1,2^k+1,2^{3k}+1\}$ where $\gcd(k,n)=1$. We present a new proof that this triple leads to a triple-error-correcting code. Furthermore, the main result in this paper is to find a new triple given by $\{1,2^k+1,2^{2k}+1\}$ for any $n$ where $\gcd(k,n)=1$.

Let $d_1=1$, $d_2=3$,  and $d_3=5$, then we can construct the parity-check matrix $H$ of the triple-error-correcting BCH code as follows:
 $$H =\left[\begin{array}{ccccc}
        1 & \alpha^{d_1}    & \alpha^{2 d_1}    & \cdots  & \alpha^{(2^n-2)d_1}      \\
        1 & \alpha^{d_2}    & \alpha^{2 d_2}    & \cdots  & \alpha^{(2^n-2)d_2}  \\
        1 & \alpha^{d_3}    & \alpha^{2 d_3}    & \cdots  & \alpha^{(2^n-2)d_3}
 \end{array} \right].$$

In general we are interested in finding triples $\{d_1,d_2,d_3\}$ such that $H$ is the parity-check matrix of a triple-error-correcting code $C$. Each column is a binary vector of length $3n$ composed of the binary representations of three elements of $GF(2^n)$ with respect to some chosen basis. This matrix is a  $3n$ by $2^n-1$ array and hence $C$ has parameters $[2^n-1, 2^n-3n-1, d]$ (except for some degenerate cases). This means that $C$ is a code of dimension $2^n-3n-1$ and minimum Hamming distance $d=7$ between any pair of codewords.

\section{Known Triple-Error-Correcting Codes}

The following table lists some of the known triples that lead to triple-error-correcting codes that can be constructed by the parity-check matrix $H$ above.

\bigskip

\begin{center}

 \begin{tabular}{|c|c|c|}
 \hline
  &  &  \\
 ${\bf Triples }$ & {\bf Conditions} & {\bf References }\\
 \hline
  &  &  \\
 \small{$\{1, 2^k+1, 2^{2k}+1\}$} & $gcd(n,k)=1$  & \\
  & any $n$ & Theorem 1 \\
 \hline
   &  & \\
 \small{$\{1, 2^k+1, 2^{3k}+1\}$} &  $gcd(n,k)=1$  & \small{\cite{Kas1}}  \\
 & $n$ odd & Theorem 2\\
 \hline
  &  &  \\
 \small{$\{1, 2^{t}+1 , 2^{t+2}+3\}$} &  $n=2t+1$  & \small{\cite{Cha}}\\
  & $n$ odd &  \\
 \hline
  &  &  \\
 \small{$\{2^k+1, 2^{3k}+1, 2^{5k}+1\}$} &   $gcd(n,k)=1$ & \small{\cite{Kas1}}   \\
  & $n$ odd &  \\
  \hline
  &  &  \\
 \small{$\{1, 2^t+1, 2^{t-1}+1\}$} &  $n=2t+1$  & \small{\cite{MS}} \\
  & $n$ odd &  \\
 \hline
 \end{tabular}.
\end{center}

\bigskip

\section{New triple-error-correcting Codes}

In this section we consider zero sets from two triples that lead to the construction of triple-error-correcting codes. First we consider the new triple $\{1,2^k+1,2^{2k}+1\}$ where $\gcd(k,n)=1$ and where $n$ can be odd or even. The $n$ even case is new while the $n$ odd case is a consequence of Kasami~\cite{Kas1}. Thereafter, we apply a similar  technique to provide a new and simpler proof of the triple $\{1,2^k+1,2^{3k}+1\}$, where $n$ is odd and $\gcd(k,n)=1$, that was shown by Kasami~\cite{Kas1} to lead to triple-error-correcting codes. Furthermore, we demonstrate that this triple will not lead to a code of distance seven when $n$ is even. We believe these results will provide further insight into the problem since it shows some new and interesting connections to the number of solutions of some special polynomials over finite fields given in Lemma~\ref{lem:Bracken} and Lemma~\ref{lem:Bluher}.

\bigskip

We shall make use of the following lemma, a proof of which can be found in Bracken et. al.~\cite{BBMM}.

\medskip

\begin{lemma} \label{lem:Bracken}
Let $s$ be an integer satisfying $\gcd(s,n)=1$ and let $f(x) = \sum_{i=0}^d r_i x^{2^{si}}$ be a polynomial in $GF(2^n)[x]$. Then $f(x)$ has at most $2^d$ zeroes in $GF(2^n)$.
\end{lemma}

\bigskip

The following lemma is a consequence of a result in Bluher~\cite{Blu}.

\medskip

\begin{lemma} \label{lem:Bluher}
An equation of the form $x^{2^k+1}+bx^{2^k}+cx=d$ defined on $GF(2^n)$, has no more than three solutions in $x$ when $gcd(k,n)=1$ for all $b, \ c,$ and $d$ in $GF(2^n)$.
\end{lemma}

\bigskip

One technique for determining the minimum distance of an error-correcting code is to use the fact that, if there are no sets of $d-1$ linearly dependent columns in $H$, then the code $C$ has distance of at least $d$. This fact is easily derived from the fact that $C$ is the nullspace of $H$. To obtain our results we will assume that $H$ has six linearly dependent columns and derive a contradiction, thus demonstrating a minimum distance of seven.

\medskip

A useful observation is that the codes with zero sets $\{1,2^k+1\}$ where $\gcd(k,n)=1$ have minimum distance five. For $k=1$ this follows since the code is the double-error-correcting BCH codes. The general result is a simple consequence of the well known result that $f(x)=x^{2^k+1}$ is an almost perfect nonlinear(APN) power functions when $\gcd(k,n)=1$. Note that an APN function is a function such that $f(x+a)+f(x) = b$ has at most two solutions $x \in GF(2^n)$ for any $a \neq 0$ and $b$ in $GF(2^n)$. For more information about APN functions the reader is referred to Bracken et. al.~\cite{BBMM} and Dobbertin~\cite{Dob}.

\medskip

\begin{theorem} \label{the:BraHel}
The error-correcting code with the zero set $\{1, 2^k+1, 2^{2k}+1 \}$ is triple-error-correcting provided $\gcd(k,n)=1$.
\end{theorem}

\medskip

{\em Proof:} If $H$ has six or less dependent columns then there must exist elements $x,y,z,u,v,w$ in $GF(2^n)$ such that,
$$ x+y+z+u+v+w=0$$
$$x^{2^k+1}+y^{2^k+1}+z^{2^k+1}+u^{2^k+1}+v^{2^k+1}+w^{2^k+1}=0$$
$$x^{2^{2k}+1}+y^{2^{2k}+1}+z^{2^{2k}+1}+u^{2^{2k}+1}+v^{2^{2k}+1}+w^{2^{2k}+1}=0,$$
has a nontrivial solution (i.e., not all being zero or pairwise equal).

Note that since the code with zero set $\{1,2^k+1\}$, where $\gcd(k,n)=1$, has minimum distance five it follows from the first two equations that all elements $x,y,z,u,v,w$ have to be distinct.

We can write this as
$$x+y+z=a=u+v+w$$
$$x^{2^k+1}+y^{2^k+1}+z^{2^k+1}=b=u^{2^k+1}+v^{2^k+1}+w^{2^k+1}$$
$$x^{2^{2k}+1}+y^{2^{2k}+1}+z^{2^{2k}+1}=c=u^{2^{2k}+1}+v^{2^{2k}+1}+w^{2^{2k}+1},$$
for some $a,b,c$ $\in GF(2^n)$.

Note that $b \neq a^{2^k+1}$ since otherwise there would be a codeword of weight 4 with error locations $\{x,y,z,a\}$, contradicting that the code with zero set $\{1,2^k+1\}$, where $\gcd(k,n)=1$, has minimum distance five.

In order to obtain the required contradiction we will demonstrate that the following system
$$x+y+z=a$$
$$x^{2^k+1}+y^{2^k+1}+z^{2^k+1}=b$$
$$x^{2^{2k}+1}+y^{2^{2k}+1}+z^{2^{2k}+1}=c,$$
cannot have six solutions in $x$ for any $a,b,c$ $\in GF(2^n)$.

Replace $x$ with $x+a$, $y$ with $y+a$ and $z$ with $z+a$. Then the system becomes,
$$x+y+z=0$$
$$(x+a)^{2^k+1}+(y+a)^{2^k+1}+(z+a)^{2^k+1}=b$$
$$(x+a)^{2^{2k}+1}+(y+a)^{2^{2k}+1}+(z+a)^{2^{2k}+1}=c.$$

Expanding the second and third equations and using the first one that $x+y+z=0$, leads to the following simple equation system.
$$ x+y+z=0$$
$$x^{2^k+1}+y^{2^k+1}+z^{2^k+1}=b+a^{2^k+1}$$
$$x^{2^{2k}+1}+y^{2^{2k}+1}+z^{2^{2k}+1}=c+a^{2^{2k}+1}.$$
Substituting $z=x+y$ leads to
$$x^{2^{k}}y+y^{2^{k}}x = \beta $$
$$x^{2^{2k}}y+y^{2^{2k}}x = \gamma $$
where $\beta=b+a^{2^k+1}$ and $\gamma = c+a^{2^{2k}+1} $. Note in particular that $\beta \neq 0$ since we already showed above that $b \neq a^{2^k+1}$. We now replace $y$ with $xy$ and get,

$$\ \ \ \ \ \ x^{2^{k}+1}(y+y^{2^{k}}) = \beta  \ \ \ \ \ \ \ \ \ \ \ \ (1)$$
$$\ \ \ \ \ \ x^{2^{2k}+1}(y+y^{2^{2k}}) = \gamma. \ \ \ \ \ \ \ \ \ \ \ (2)$$
A rearrangement of Equation (1) yields
$$y+y^{2^{k}}=\beta x^{-2^{k}-1},$$
which implies
$$y+y^{2^{2k}}=\beta x^{-2^{k}-1}+{\beta}^{2^k} x^{-2^{2k}-2^k}.$$
We can now place this expression for $y+y^{2^{2k}}$ into Equation (2) and obtain
$$x^{2^{2k}+1} (\beta x^{-2^{k}-1}+{\beta}^{2^k} x^{-2^{2k}-2^k})= \gamma.$$
This becomes the linearized equation
$$\beta x^{2^{2k}} + {\beta}^{2^k}x + \gamma x^{2^k}=0,$$
which since $\beta \neq 0$ has no more than four solutions in $x$ by Lemma~\ref{lem:Bracken} and we are done.
\hspace*{\fill}${\Box}$

\bigskip

The following result was first proved by Kasami~\cite{Kas1} but we provide another proof as it is rather elementary and we believe it provides further insight into the problem. In particular the proof shows an interesting connection to the equation in Lemma~\ref{lem:Bluher}.

\medskip

\begin{theorem} \label{the:Kas}
Let $n$ be odd and $\gcd(k,n)=1$. Then the error-correcting code constructed using the zero set $\{1, 2^k+1, 2^{3k}+1\}$ is triple-error-correcting.
\end{theorem}

\medskip

{\em Proof:}
Using the same arguments as in Theorem~\ref{the:BraHel} we obtain the pair of equations
$$\ \ \ \ \ \ x^{2^{k}+1}(y+y^{2^{k}}) = \beta  \ \ \ \ \ \ \ \ \ \ \ \ (1)$$
$$\ \ \ \ \ \ x^{2^{3k}+1}(y+y^{2^{3k}}) = \gamma, \ \ \ \ \ \ \ \ \ \ \ (2)$$
where $\beta=b+a^{2^k+1} \neq 0$ and $\gamma = c+a^{2^{3k}+1}$.
We rearrange Equation (1) and obtain
$$y+y^{2^{k}}=\beta x^{-2^{k}-1},$$
which implies
$$y+y^{2^{3k}}=\beta x^{-2^{k}-1}+{\beta}^{2^k} x^{-2^{2k}-2^k} +{\beta}^{2^{2k}} x^{-2^{3k}-2^{2k}} .$$
Combining this with Equation (2) we get
$$x^{2^{3k}+1}(\beta x^{-2^{k}-1}+{\beta}^{2^k} x^{-2^{2k}-2^k} +{\beta}^{2^{2k}} x^{-2^{3k}-2^{2k}})=\gamma.$$
This becomes
$$\beta x^{2^{3k}-2^{k}}+{\beta}^{2^k} x^{2^{3k}+1-2^{2k}-2^k} +{\beta}^{2^{2k}}x^{1-2^{2k}}=\gamma.$$
Next let $r=x^{2^{2k}-1}$. This substitution is one-to-one as $n$ is odd.
We now have
$$\beta r^{2^k}+{\beta}^{2^k} r^{2^{k}-1} +{\beta}^{2^{2k}}r^{-1}=\gamma,$$
which implies
$$\beta r^{2^k+1}+{\beta}^{2^k} r^{2^{k}}+\gamma r +{\beta}^{2^{2k}} =0.$$
Since $\beta \neq 0$ it follows by Lemma 2 that this equation can have no more than three solutions in $r$. Hence, we conclude that the code has minimum distance at least seven and we are done.
\hspace*{\fill}${\Box}$

\bigskip
It is interesting to observe that in Theorem~\ref{the:Kas} we need $n$ to be odd (and $\gcd(k,n)=1$) while this is not needed in Theorem~\ref{the:BraHel}. We will show that this condition is necessary in order for Theorem~\ref{the:Kas} to hold.

\medskip
Let $n$ is even and we will show that the code with zero set corresponding to the triple $\{1, 2^k+1, 2^{3k}+1\}$ has minimum distance at most six. We will show this by constructing a nonzero codeword of Hamming weight at most six.

\medskip

{\em Case 1:} Let $k$  be odd (and $n$ even). Let $\delta$ be a primitive element in $GF(2^2) \subset GF(2^n)$ and thus $\delta^3=1$. Let $x, y$ and $z$ be selected as nonzero elements in $GF(2^n)$ such that $x+y+z=0$ and $x/y \not \in \{1,\delta,\delta^2\}$. Then $\{x,y,z,\delta x,\delta y,\delta z\}$ are the locations for a nonzero codeword of weight six. This follows since the conditions imply that the locations are all distinct and $k$ is odd and thus $\delta^{2^k+1} = \delta^{2^{3k}+1} = 1$ and therefore
\[ x + y + z =  \delta x + \delta y + \delta z \]
\[ x^{2^k+1} + y^{2^k+1} + z^{2^k+1} = (\delta x)^{2^k+1} + (\delta y)^{2^k+1} + (\delta z)^{2^k+1} \]
\[ x^{2^{3k}+1} + y^{2^{3k}+1} + z^{2^{3k}+1} = (\delta x)^{2^{3k}+1} + (
\delta y)^{2^{3k}+1} + (\delta z)^{2^{3k}+1}. \]

\medskip

{\em Case 2:} Let $k$ (and $n$) be even. Let $\delta$ be a primitive element in $GF(2^2) \subset GF(2^n)$.  Select $x$ to be a nonzero element in $GF(2^n)$. Then there is a codeword in the code of weight three with error locations $x, x \delta$ and $x \delta^2$.

This follows since the conditions imply that the locations are all distinct and since $k$ is even we have  $\delta^{2^k+1} + \delta^{2(2^k+1)} = \delta^{2^{3k}+1} + \delta^{2(2^{3k}+1)} = 1$. Therefore it holds that
\[ x + x \delta + z \delta^2 = x(1+ \delta + \delta^2) = 0\]
\[ x^{2^k+1} + x^{2^k+1} \delta^{2^k+1} + x^{2^k+1}\delta^{2(2^k+1)} = 0 \]
\[ x^{2^{3k}+1} + x^{2^{3k}+1} \delta^{2^{3k}+1} + x^{2^{3k}+1}\delta^{2(2^{3k}+1)} = 0. \]

\medskip

Finally we note that Theorem~\ref{the:BraHel} works for any $n$ when $\gcd(k,n)=1$. Thus this implies that when $n$ is even then $k$ must be odd. In the case when $n$ is even and $k$ is even the result will not hold since we can use the  same technique as in Case 2 to obtain a codeword of weight three.
 
\section{Conclusion}

A new triple-error-correcting code has been constructed and a new simpler proof of one of the triple-error-correcting codes by Kasami has been presented. The proofs show some new connections to properties of some special equations of finite fields given in Lemma~\ref{lem:Bracken} and Lemma~\ref{lem:Bluher}. Finding further triples leading to new triple-error-correcting codes is a fascinating and challenging research problem that may lead to other interesting connections.

\section*{Acknowledgment}

This research was supported by the Norwegian Research Council.

\end{document}